\documentclass[prd, 10pt,aps,nofootinbib, floatfix, notitlepage,twocolumn,longbibliography]{revtex4-1}

\usepackage{amsmath,amsfonts,amssymb,mathrsfs} 
\usepackage{graphicx}
\usepackage{dcolumn}
\usepackage{array}
\usepackage{bm}
\usepackage{xcolor}

\PassOptionsToPackage{colorlinks=true,citecolor=blue,urlcolor=blue}{hyperref}
\usepackage{orcidlink}
\usepackage[english]{babel}
\usepackage{url}
\usepackage{adjustbox}
\usepackage[utf8]{inputenc}

\def\bs{\boldsymbol}
\newcommand{\code}[1]{\texttt{#1}}

\newcommand{\cph}[1]{\textcolor{red}{[cite]}}
\begin{document}

\title{Lowering the Horizon on Dark Energy: A Late-Time Response to Early Solutions for the Hubble Tension}
\author{Tal Adi\;\orcidlink{0000-0002-5763-9353}}
\email{taladi@usc.edu}
\affiliation{University of Southern California, Los Angeles, CA 90089, USA}

\begin{abstract}
We present a model-independent null test of the late-time cosmological response to a reduced sound horizon, as typically required by early-universe solutions to the Hubble tension. In this approach, we phenomenologically impose a shorter sound horizon without modeling early-universe physics to isolate its impact on late-time dark energy inference. Using baryon acoustic oscillations (BAO), supernovae (SN), big bang nucleosynthesis (BBN), and local $H_0$ data, while explicitly avoiding CMB anisotropies, we examine how this calibration shift propagates into constraints on the dark energy equation of state. We find that lowering $r_d$ systematically drives the $w_0$-$w_a$ posterior toward less dynamical, quintessence-like behavior, bringing it closer to $\Lambda$CDM. This result underscores that some of the apparent evidence for evolving or phantom-like dark energy may reflect early-universe assumptions rather than genuine late-time dynamics. More broadly, our analysis highlights the importance of carefully disentangling calibration effects from physical evolution in interpreting forthcoming results from DESI and future surveys.
\end{abstract}

\maketitle
 
\section{Introduction}

One of the central challenges in modern cosmology is understanding the nature of dark energy (DE) and its role in driving the accelerated expansion of the universe~\cite{Weinberg:2013agg,Martin:2012bt,Burgess:2013ara,Padilla:2015aaa}. Recent results from the Dark Energy Spectroscopic Instrument (DESI) have offered new insights into DE evolution, showing a mild preference for a dynamical, phantom-like equation of state at redshifts $z \gtrsim 0.5$~\cite{DESI:2025zgx,DESI:2025wyn,DESI:2025fii}. If confirmed, this could mark a significant departure from the cosmological constant paradigm, prompting renewed theoretical interest~\cite{Chen:2025ywv,Yao:2025wlx,Nojiri:2025low,Pan:2025psn,Linder:2025zxb,Goldstein:2025epp,Khoury:2025txd} and potentially pointing to new physics beyond $\Lambda$CDM~\cite{VanWaerbeke:2025shm}.

A parallel and possibly related tension arises in the measurement of the Hubble constant $H_0$\cite{DiValentino:2021izs,Verde:2023lmm}. Local determinations using the distance ladder report $H_0 = 73.04 \pm 1.04$ km/s/Mpc~\cite{Riess:2021jrx}, in significant disagreement with the value inferred from Planck CMB measurements under $\Lambda$CDM, $H_0 = 67.4 \pm 0.5$ km/s/Mpc~\cite{Planck:2018vyg}. While efforts are ongoing to rule out systematics~\cite{Riess:2021jrx,Abdalla:2022yfr,Breuval:2024lsv,CosmoVerseNetwork:2025alb}, the persistence and significance of this discrepancy—known as the Hubble tension—suggest it may be a genuine hint of new physics.

Proposed solutions~\cite{Schoneberg:2021qvd} to the Hubble tension typically fall into two categories: modifications to early-universe physics and late-time extensions to the DE sector~\cite{Knox:2019rjx}. The former class includes models such as Early Dark Energy (EDE)~\cite{Karwal:2016vyq,Poulin:2018cxd,Poulin:2018dzj,Smith:2019ihp}, primordial magnetic fields~\cite{Jedamzik:2020krr,Jedamzik:2023csc,Jedamzik:2025cax}, early modified gravity~\cite{Rossi:2019lgt,Braglia:2020iik,Braglia:2020auw,Zumalacarregui:2020cjh,Adi:2020qqf,Ballardini:2020iws,FrancoAbellan:2023gec}, and others~\cite{Berghaus:2019cls,Agrawal:2019lmo,Niedermann:2020dwg,Lin:2020jcb,Ye:2020btb,Ballesteros:2020sik,Dwivedi:2024okk,Hogas:2025ahb}. A common feature of many early-time solutions is a reduction in the sound horizon, $r_s$—the maximum comoving distance that acoustic waves could travel in the primordial plasma before decoupling. This scale serves as a standard ruler in both CMB and BAO observations. A shorter $r_s$ effectively rescales inferred cosmological distances and drives a higher value of $H_0$.

If the sound horizon $r_s$ is shorter than currently estimated, it would directly affect late-time DE inference. Given that $r_s$ calibrates the acoustic oscillation scale observed in the CMB and large-scale structure, a smaller value of $r_s$ implies a different expansion history and could alter the inference of DE equation-of-state (EoS). However, any early-universe mechanism invoked to shorten $r_s$ will inevitably also shift other cosmological parameters to maintain consistency with different data. For example, an EDE scenario that reduces the sound horizon typically also alters the spectral index $n_s$ and the cold dark matter density $\omega_{\rm cdm}$~\cite{Hill:2020osr,Vagnozzi:2021gjh,Vagnozzi:2023nrq,Poulin:2025nfb}. This interplay makes it difficult to attribute changes in DE constraints solely to a smaller sound horizon, highlighting the need to assess this effect in a model-independent way.
More broadly, the calibrator picture\cite{Aylor:2018drw,Poulin:2024ken,Pedrotti:2024kpn} shows that simply shrinking $r_s$ through early-universe changes is not sufficient: the model must also account for the larger $\omega_m$. Otherwise, both early- and late-time modifications are needed.

In this work, we perform a model-independent\footnote{Here we mean that we do not assume a specific early-Universe solution to the Hubble tension; however, we do assume a CPL parametrization for dark energy.} null test to isolate the late-time response of DE to a reduced sound horizon, as typically required by early-time solutions to the Hubble tension. We do not propose a solution to the Hubble tension; we impose the reduced sound horizon phenomenologically to test how such early-time fixes would propagate into late-time CPL constraints.

\section{Methodology}

\begin{figure*}[ht!]
    \centering
    \includegraphics[width=0.45\textwidth]{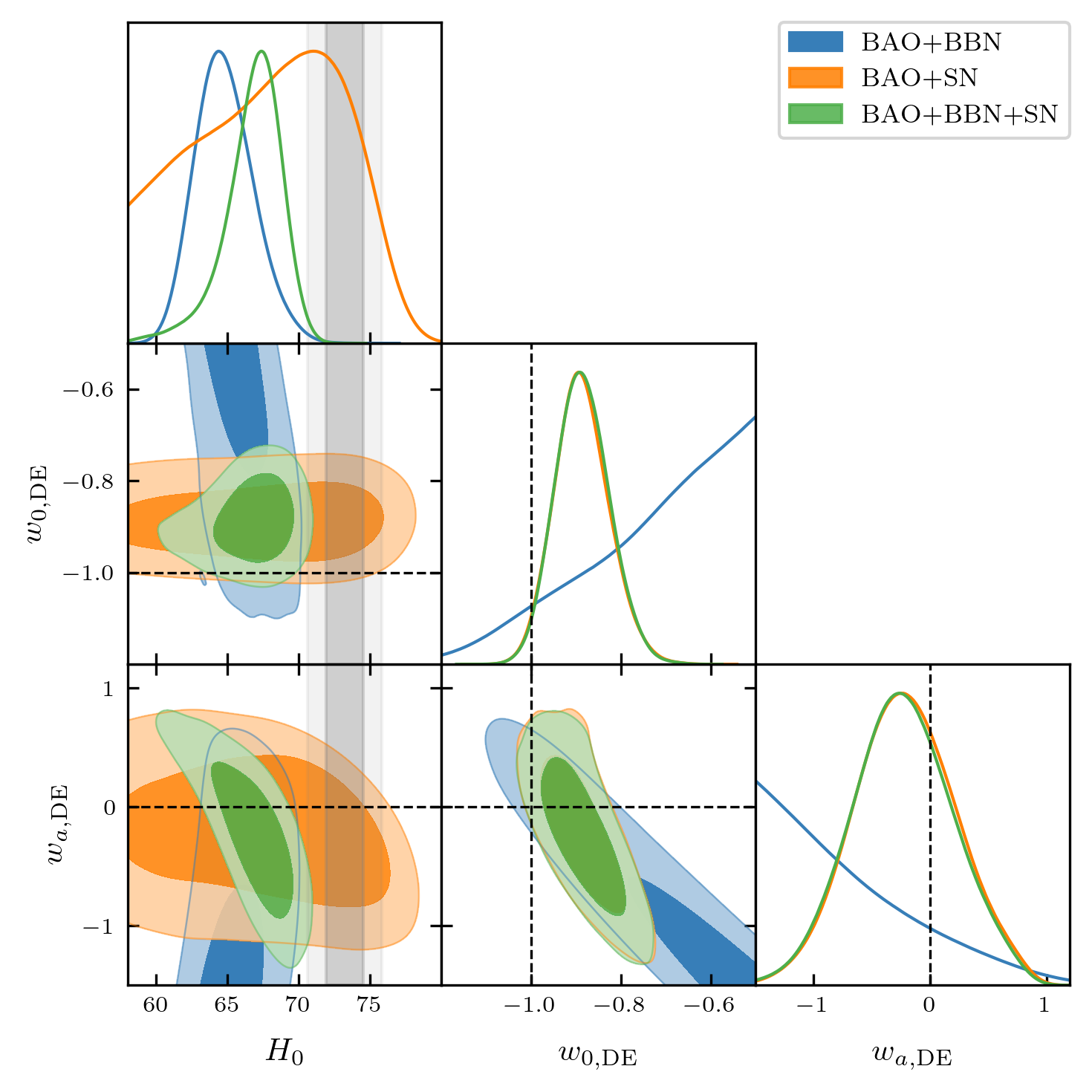}
    \includegraphics[width=0.45\textwidth]{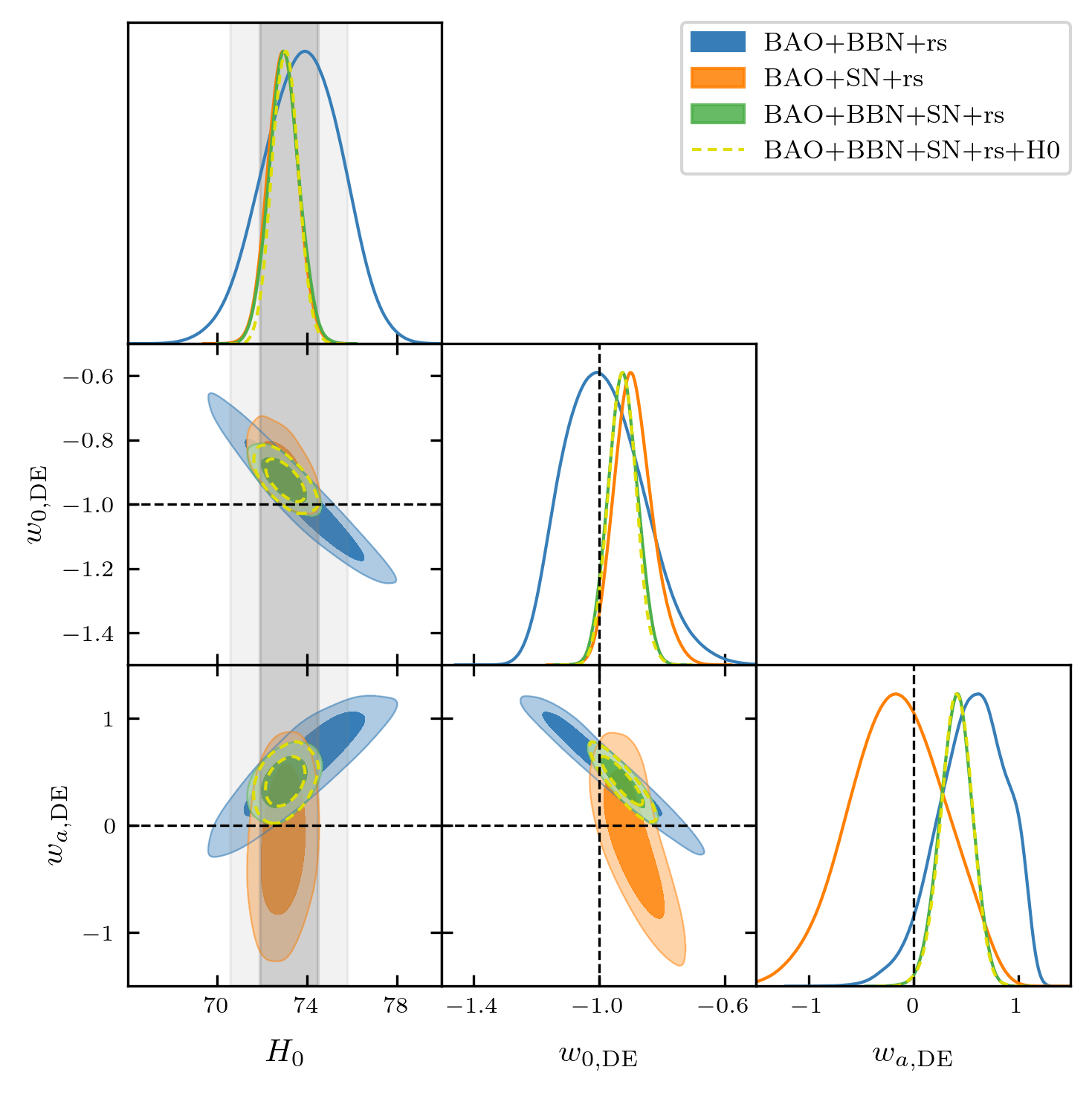}
    \caption{Posterior distributions of cosmological parameters with (right panel) and without (left panel) the $r_d$ prior. Each triangle plot shows the 68\% and 95\% confidence regions for key parameters using the baseline combinations of BAO, BBN, and SN data. Shaded bands indicate the SH0ES~\cite{Riess:2021jrx} confidence regions for $H_0$, and black dashed lines guide the eye to $w_0 = -1$ and $w_a = 0$.
    The inclusion of a reduced sound horizon prior shifts $H_0$ to higher values and induces correlated changes in the CPL parameters ($w_0$, $w_a$), reflecting the late-time response required to maintain consistency with BAO and SN observations.}
    \label{fig:triangle}
\end{figure*}

To parametrically explore deviations from a cosmological constant ($w = -1$), we adopt the widely used Chevallier-Polarski-Linder (CPL) parametrization for the DE EoS:
\begin{equation}\label{eq:cpl}
    w(z) = w_0 + w_a \frac{z}{1+z},
\end{equation}
where $w_0$ is the present-day value of the EoS and $w_a$ characterizes its evolution with scale factor $a = 1/(1+z)$. This parametrization, effectively a first-order Taylor expansion around $a = 1$, captures a wide range of dynamical DE models~\cite{Chevallier:2000qy,Linder:2002et,Linder:2024rdj} including quintessence ($-1<w<-1/3$)~\cite{Frieman:1995pm,Shajib:2025tpd} and phantom DE ($w<-1$)~\cite{Caldwell:1999ew}. The DESI DR2 results~\cite{DESI:2025zgx}, when combined with uncalibrated supernova (SN) distances and CMB data, show a clear preference for deviations from $w = -1$, with $w_0 \approx -0.75 \pm 0.067$ and $w_a \approx -0.86 \pm 0.25$, depending on the SN sample. This suggests a DE scenario that is transitioning from phantom to quintessence around $z \approx 0.5$, implying a weakening of the cosmic acceleration at lower redshifts.

The sound horizon at redshift $z$ is given by:
\begin{equation}\label{eq:rs}
    r_s(z) = \int_{z}^\infty \frac{c_s(z')}{H(z')} dz',
\end{equation}
where $c_s(z)$ is the sound speed of the photon-baryon plasma. This can be evaluated either at recombination ($z_\mathrm{rec} \sim 1090$) or at the baryon drag epoch ($z_d \sim 1060$)~\cite{Planck:2018vyg}. We focus on the sound horizon at $z_d$, denoted $r_d$, as it is the relevant scale for large-scale structure and BAO observations~\cite{Eisenstein:1997ik}.

Models that reduce $r_d$ typically do so by increasing the expansion rate $H(z)$ prior to recombination, which can be achieved by adding new energy components or modifying gravity. However, a phenomenological approach to reducing the sound horizon, without accounting for early-time modifications, will not be consistent when considering CMB data. First and foremost, this is due to the precise measurement of the angular scale of the sound horizon at recombination, $\theta_s = r_s/D_A(z_{\rm rec})$, where $D_A(z_{\rm rec})$ is the angular diameter distance to the last scattering surface~\cite{Hu:1996qz,Pan:2016zla}. Lowering $r_s$ would lead to a shift in the cosmological ($\Lambda$CDM) parameters, that are known to be incapable of accommodating a higher $H_0$~\cite{Knox:2019rjx}.

To avoid these complications and isolate the late-time response, we do not include CMB anisotropy data. Instead, we restrict our analysis to late-time probes and sound-horizon-independent observations that help break parameter degeneracies.

At low redshifts, the Hubble parameter for a flat universe is:
\begin{equation}\label{eq:hubble}
    H(z) = H_0 \sqrt{\Omega_{m}\left(1+z\right)^{3}+\Omega_{\text{DE}}f\left(z\right)},
\end{equation}
where $\Omega_m$ and $\Omega_{\rm DE}$ are the present-day matter and DE densities, and
\begin{equation}
    f(z) = \left(z+1\right)^{3\left(w_{0}+w_{a}+1\right)}e^{-3w_{a}z/\left(z+1\right)}
\end{equation}
describes the evolution of DE from the CPL parametrization~\eqref{eq:cpl}.

We use the Boltzmann code \code{CLASS}\footnote{\url{https://github.com/lesgourg/class_public}}~\cite{Blas:2011rf} to compute the background evolution and linear perturbations, and sample the cosmological parameters $\{w_0, w_a, H_0, \omega_b, \omega_\mathrm{cdm}\}$ using MCMC as implemented in \code{Cobaya}\footnote{\url{https://github.com/CobayaSampler/cobaya}}~\cite{Torrado:2020dgo}. We adopt flat priors for all parameters as in \cite{DESI:2025zgx,DESI:2024mwx}, including a hard prior on $w\left(z\gg1\right)<0$.
To decouple early-universe physics from our analysis, we fix $\{A_s, n_s, \tau_\mathrm{reio}\}$ to their Planck 2018 best-fit values~\cite{Planck:2018vyg}, and verify that varying them does not significantly affect our results. We use the Gelman-Rubing convergence criterion~\cite{Gelman:1992zz} $R-1 < 0.01$ to ensure robust sampling, and analyze the chains using \code{GetDist}\footnote{\url{https://github.com/cmbant/getdist}}~\cite{Lewis:2019xzd}.

\subsection*{Datasets}

\textbf{BAO measurements from DESI DR2} (BAO): BAO measurements provide a direct probe of the expansion history of the universe by measuring the excess of clustering in the galaxy distribution at the scale of the sound horizon~\cite{McDonald:2006qs}. Specifically, they provide constraints on both the transverse and radial distances, encoded via:
\begin{align}
    \frac{D_M(z)}{r_d} = \frac{1}{r_d}\int_0^z \frac{dz'}{H(z')},\label{eq:dm} \\
    \frac{D_H(z)}{r_d} = \frac{1}{r_d H(z)},\label{eq:dh}
\end{align}
where $D_M(z)$ is the \emph{comoving} angular diameter distance (transverse), and $D_H(z)$ is the Hubble distance (radial). We use BAO data from galaxies and quasars spanning $0.2 < z < 3.5$~\cite{DESI:2025zgx,DESI:2025zpo}.

\textbf{Type Ia supernovae} (SN): These are standard candles that probe the late-time expansion at low redshifts ($0.01 < z < 0.3$), complementing BAO measurements. We include three datasets: Pantheon+~\cite{Brout:2022vxf}, Union3~\cite{Rubin:2023ovl}, and Dark Energy Survey Year 5 (DESY5)~\cite{DES:2024jxu}. To avoid the clutter, throughout this study, we use Pantheon+ as a representative dataset; we confirm that substituting Union3 or DESY5 does not alter our main conclusions.

\textbf{Big Bang Nucleosynthesis} (BBN): The primordial abundances of light elements, such as deuterium and helium, are sensitive to the physical baryon density $\omega_b = \Omega_b h^2$ and the expansion rate during BBN. We use the latest measurements of primordial deuterium abundance from quasar absorption systems~\cite{Schoneberg:2024ifp}, which provide a precise constraint on $\omega_b$. BBN constraints are independent of the sound horizon and provide a useful prior on $\omega_b$, which is degenerate with $H_0$ in late-time observations. We use a Gaussian prior of $\omega_b = 0.0222 \pm 0.0005$. While BBN constraints are largely orthogonal to late-time probes and independent of $r_d$, certain early-universe mechanisms (e.g., varying $N_\text{eff}$~\cite{Schoneberg:2024ifp}) can shift the BBN-inferred $\omega_b$. Our analysis accounts for this by confirming stability of results under modest variations in the $\omega_b$ prior.

\textbf{Local $H_0$ measurements} (H0): We also consider the local distance ladder measurement of $H_0$ from SH0ES~\cite{Riess:2020fzl}, which provides a direct measurement of the current expansion rate of the universe. This measurement is independent of the sound horizon and can help break degeneracies in late-time observations. We adopt a Gaussian prior of $H_0 = 73.04 \pm 1.04 \, \text{km/s/Mpc}$~\cite{Riess:2021jrx}.

\textbf{Sound horizon prior} (rs): To mimic the effect of early-time models that reduce $r_d$, we impose a Gaussian prior of $r_d = 136.8 \pm 0.24$, based on the value typically required to alleviate the Hubble tension~\cite{Knox:2019rjx}.
We implement the shorter sound horizon prior phenomenologically by rescaling the Planck 2018 best-fit value $r_d^\mathit{Planck} = 147.1$~\cite{Planck:2018vyg} by a factor of $0.93$\footnote{While we adopt $r_d = 0.93\,r_d^\mathit{Planck}$ as a representative calibration, Section~\ref{sec:results_and_discussion} explicitly explores the dependence of the inferred parameters on the imposed sound horizon.} within the BAO likelihood, yielding $r_d = 136.8$~Mpc. This approach ensures internal consistency of the BAO likelihood without altering early-universe physics directly. By construction, this approach does not enforce consistency with CMB anisotropies, but instead serves as a deliberate null test to isolate the late-time response of dark energy parameters.

\begin{figure}
    \centering
    \includegraphics[width=1\columnwidth]{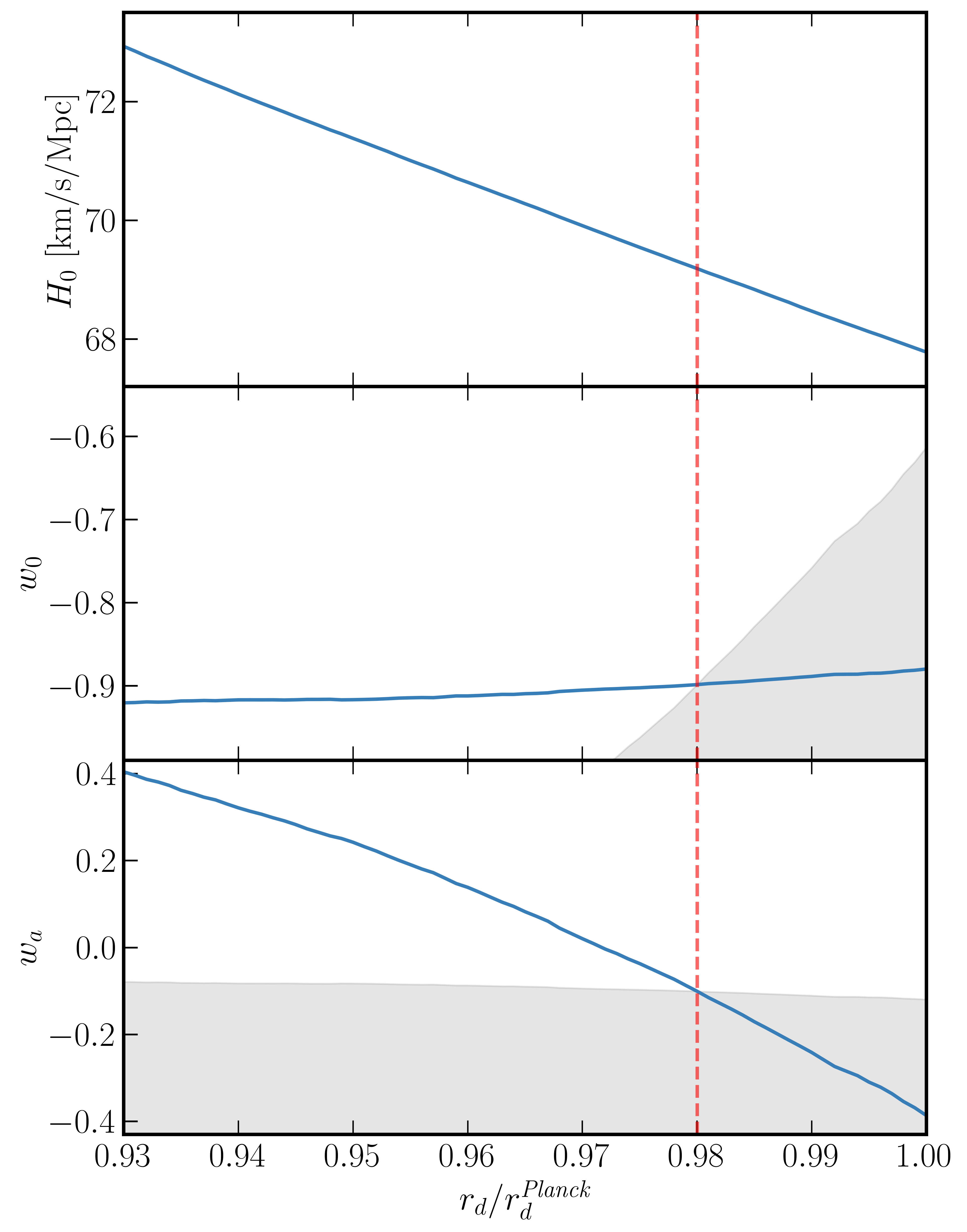}
    \caption{Evolution of the best-fit values of $H_0$, $w_0$, and $w_a$ as functions of the imposed sound horizon $r_d$, normalized to the Planck 2018 best-fit value. The results are shown for the BAO+BBN+SN+rs dataset combination. The shaded regions indicate the phantom regime ($w\le-1$), while the red dashed line marks the crossing of the phantom divide.}
    \label{fig:var_rd_w0_wa_H0}
\end{figure}

Rescaling $r_d$ without modifying the early-universe physics is internally inconsistent with CMB anisotropy measurements. Therefore, our approach should be viewed as a null test designed to isolate the late-time response of DE parameters; we do not claim that this setup represents a viable cosmological model.

\section{Results and Discussion}\label{sec:results_and_discussion}

\begin{table*}
\footnotesize
\renewcommand{\arraystretch}{1.5}
    \setlength{\tabcolsep}{5pt}
    \begin{adjustbox}{max width=\textwidth}\begin{tabular}{l | c c c c c c c}
        \hline\hline
        & \multicolumn{7}{c}{BAO +} \\
        Parameter & BBN & BBN+rs & SN & SN+rs & BBN+SN & BBN+SN+rs & BBN+SN+rs+H0 \\
        \hline\hline
        $\bs{H_0}$ & $64.8^{+1.8}_{-2.22}$ & $73.78^{+1.82}_{-1.79}$ & $66.73^{+7.95}_{-4.66}$ & $72.91^{+0.66}_{-0.65}$ & $66.75^{+2.3}_{-1.37}$ & $72.93^{+0.67}_{-0.65}$ & $72.99^{+0.59}_{-0.58}$ \\
        $100\bs{\omega_b}$ & $2.221\pm0.05$ & $2.211\pm0.05$ & $2.243^{+1.024}_{-0.383}$ & $1.793^{+0.12}_{-0.339}$ & $2.221\pm0.05$ & $2.209^{+0.051}_{-0.049}$ & $2.209^{+0.052}_{-0.049}$ \\
        $\bs{\omega_{cdm}}$ & $0.1255^{+0.0122}_{-0.0068}$ & $0.1209\pm0.0025$ & $0.1128^{+0.0235}_{-0.0228}$ & $0.1401^{+0.0156}_{-0.0065}$ & $0.1126^{+0.0175}_{-0.0099}$ & $0.1212^{+0.0023}_{-0.0024}$ & $0.1212^{+0.0023}_{-0.0024}$ \\
        $\bs{w_0}$ & $-0.47^{+0.33}_{-0.17}$ & $-0.98^{+0.11}_{-0.15}$ & $-0.89\pm0.06$ & $-0.89^{+0.05}_{-0.06}$ & $-0.89\pm0.06$ & $-0.92^{+0.04}_{-0.05}$ & $-0.93\pm0.04$ \\
        $\bs{w_a}$ & $-1.71^{+0.34}_{-1.29}$ & $0.55^{+0.38}_{-0.28}$ & $-0.22^{+0.45}_{-0.44}$ & $-0.15^{+0.47}_{-0.44}$ & $-0.23^{+0.45}_{-0.44}$ & $0.4^{+0.17}_{-0.14}$ & $0.4^{+0.17}_{-0.14}$ \\
        \hline
        $\Omega_m$ & $0.354^{+0.038}_{-0.02}$ & $0.264^{+0.013}_{-0.015}$ & $0.302^{+0.022}_{-0.011}$ & $0.299^{+0.024}_{-0.012}$ & $0.303^{+0.021}_{-0.011}$ & $0.271^{+0.005}_{-0.006}$ & $0.27^{+0.005}_{-0.006}$ \\
        $r_d^{\rm eff}$ & $146.0^{+1.8}_{-3.4}$ & $136.8\pm0.2$ & $150.7^{+8.1}_{-18.6}$ & $136.8\pm0.2$ & $149.5^{+2.5}_{-5.2}$ & $136.8\pm0.2$ & $136.8\pm0.2$ \\
        $\chi^2_{\rm BAO}$ & $5.63$ & $10.61$ & $8.71$ & $8.75$ & $8.69$ & $10.79$ & $10.78$ \\
        $\chi^2_{\rm SN}$ & - & - & $1402.58$ & $1402.54$ & $1402.59$ & $1402.68$ & $1402.65$ \\
        \hline
        \hline
    \end{tabular}
    \end{adjustbox}
    \caption{Constraints from BAO combined with additional datasets, showing the mean $\pm\; 1\sigma$ of sampled (bold) and derived parameters, including $\chi^2$ values for BAO and SN datasets. The columns are arranged in pairs with and without the $r_d$ prior, with the last column including $H_0$ data. $r_d^{\rm eff}$ is the effective sound horizon, defined as the rescaled $r_d$ if the $r_d$ prior is applied.}
    \label{tab:results}
\end{table*}

We begin with baseline dataset combinations (BAO+BBN, BAO+SN, BAO+BBN+SN), without any $r_d$ prior, to assess the uncalibrated constraints. The results for these cases are shown in the left panel of Fig.~\ref{fig:triangle} and summarized in Table~\ref{tab:results}.
As expected, the BAO+BBN combination provides reasonable constraints on $H_0$ but weak constraints on $w_0$ and $w_a$. This is due to the partial breaking of degeneracy between $H_0$ and $\Omega_m$ in the BAO data when the BBN prior on $\omega_b$ is applied. However, without low-redshift information, and given that BAO partially depends on integrated quantities, this combination alone does not provide sufficient constraining power for the CPL parameters. Nonetheless, the resulting posteriors remain broadly consistent with phantom DE, as suggested by Ref.~\cite{DESI:2025zgx}.
In contrast, the BAO+SN combination constrains the shape of $H(z)$ more effectively and breaks the $w_0$-$w_a$ degeneracy. However, it lacks sensitivity to the absolute scale—either through the absolute SN magnitude or a sound horizon calibration—so $H_0$ remains unconstrained. The inclusion of SN data pulls the $w_0$ and $w_a$ values closer to $-1$ and $0$, respectively, compared to BAO+BBN, albeit with large uncertainties.
The combination BAO+BBN+SN provides the most balanced constraints across all parameters. It leverages the strengths of each dataset to break degeneracies, particularly between $H_0$ and $\omega_b$, and between $w_0$ and $w_a$. The resulting $w_0$-$w_a$ posteriors are largely consistent with those from BAO+SN alone, confirming that SN dominates the constraints on the late-time expansion history. This is also evident in the minimal change in $\chi^2_\mathrm{SN}$ when adding BAO or BBN data, while $\chi^2_\mathrm{BAO}$ increases slightly.
All three baseline combinations prefer an $H_0$ value lower than the SH0ES measurement and a sound horizon in agreement with the Planck 2018 $\Lambda$CDM fit~\cite{Planck:2018vyg}. Among them, only the BAO+BBN case (without SN) favors higher $w_0$ and lower $w_a$, with a correspondingly reduced $\chi^2_\mathrm{BAO}$. This behavior reflects the known mild tension between BAO and SN datasets~\cite{Tutusaus:2023cms,Raveri:2023zmr,Bousis:2024rnb}, but remains fully consistent with the DESI DR2 results~\cite{DESI:2025zgx}.

\begin{figure}
    \centering
    \includegraphics[width=1\columnwidth]{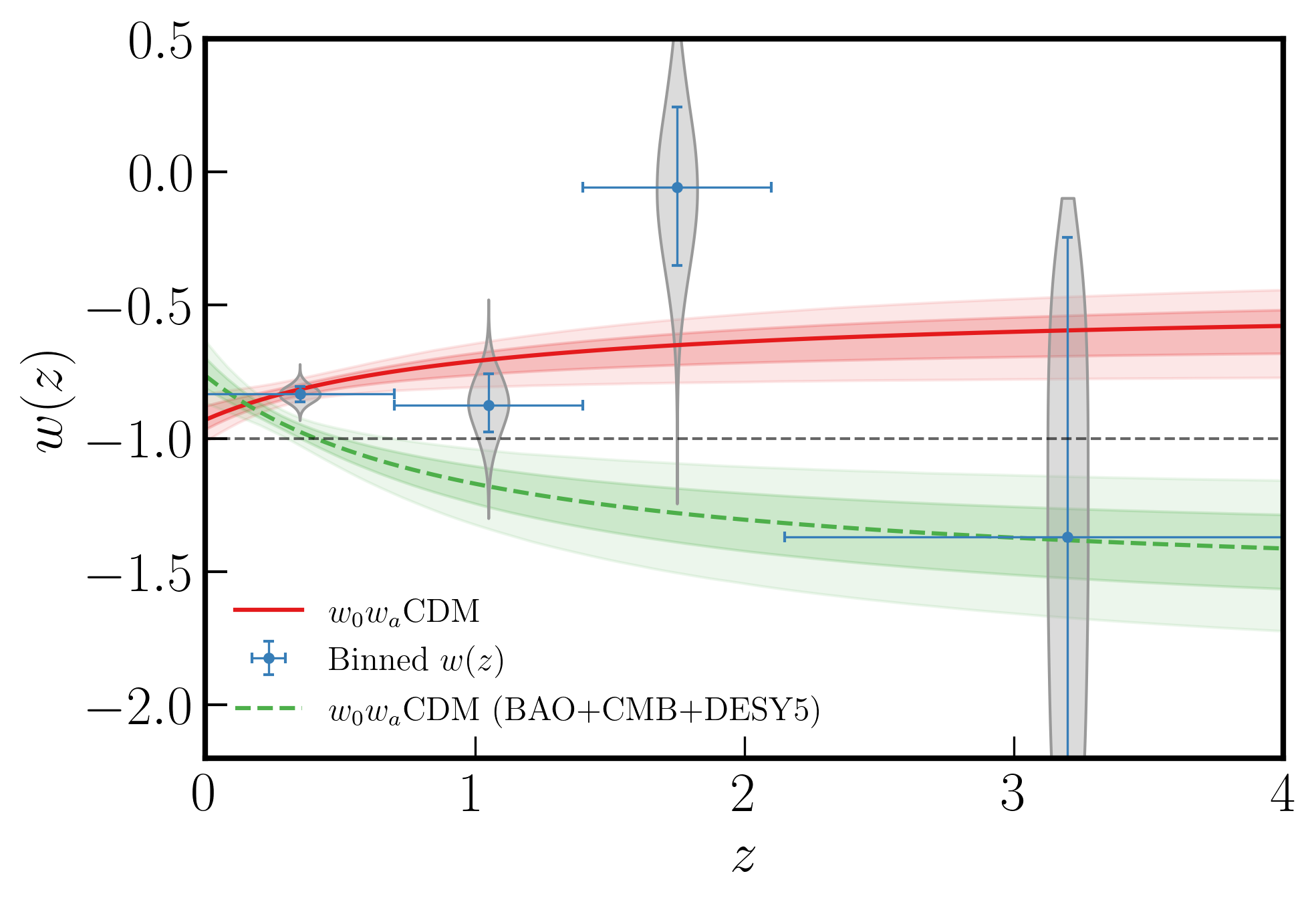}
    \caption{Comparison of DE EoS constraints obtained with the CPL parameterization and a binning reconstruction approach~\cite{DESI:2025fii}, using BAO+BBN+SN+rs data. The solid red curve indicates the best-fit $w(z)$ derived from $w_0$ and $w_a$, with shaded regions denoting the $1\sigma$ and $2\sigma$ uncertainties. Results from the binning reconstruction are shown in blue: the horizontal bars correspond to the fixed bin size and the vertical bars correspond to $1\sigma$ uncertainties. Gray contours represent the 1D posterior for the binned parameters. The green dashed curve and band show the CPL fit using the DESI DR2 dataset (BAO+CMB+DESY5)~\cite{DESI:2025zgx}, reproduced for comparison.  The horizontal gray dashed line denotes the $\Lambda$CDM expectation.}
    \label{fig:wz_binned}
\end{figure}

Imposing a lower prior on $r_d$ effectively calibrates the absolute distance scale in BAO measurements and drives $H_0$ to a higher value, as shown in the right panel of Fig.~\ref{fig:triangle} and in Table~\ref{tab:results}. This is consistent with expectations: a shorter sound horizon is often invoked to alleviate the Hubble tension.
Importantly, our analysis does not include any data sensitive to the angular acoustic scale $\theta_s$, so the observed shifts in parameters are driven entirely by late-time observations. Increasing $H_0$ effectively rescales $H(z)$, Eq.~\eqref{eq:hubble}, and to preserve consistency with BAO data, the shape of $H(z)$ must adjust. This necessitates correlated shifts in the CPL parameters: $w_0$ becomes more negative, and $w_a$ more positive. These changes reflect the need for a softer late-time evolution of DE to accommodate the elevated $H(z)$ values at low redshift.
Consistent trends appear in modified recombination and EDE scenarios, which also lower $r_d$ and reduce phantom-like dark energy~\cite{Mirpoorian:2025rfp,Wang:2024dka}.
The largest increase in $H_0$ occurs for the BAO+BBN+$r_s$ combination, where the $r_d$ prior is most effective due to the absence of SN constraints, resulting in a value that slightly overshoots the SH0ES mean. Meanwhile, $\Omega_m$ decreases slightly to accommodate higher $\Omega_\mathrm{DE}$, enhancing the effects of $w_0$ and $w_a$. We also find that $\omega_b$ is slightly reduced, while $\omega_\mathrm{cdm}$ increases, consistent with Eq.~\eqref{eq:rs}: since $c_s \propto 1/\sqrt{1 + R}$ and $R \propto \omega_b$, reducing $\omega_b$ increases $c_s$, partially compensating for the increased $H(z)$. This is corroborated by the BAO+SN and BAO+BBN+SN combinations, which prefer a lower $\omega_b/\omega_\mathrm{cdm}$ ratio when BBN is not included.
Additionally, we find that the combinations with and without SN appear to be mostly in agreement when applying the prior on $r_d$, as including it strengthens the constraints and hardly shifts the values of most parameters.

To assess whether the correlated shifts in $H_0$, $w_0$, and $w_a$ depend on a particular choice of sound-horizon rescaling, we repeat the analysis while continuously varying the imposed sound horizon over the range $0.93 \le r_d/r_d^{\rm Planck} \le 1.00$. Figure~\ref{fig:var_rd_w0_wa_H0} shows the resulting evolution of the inferred parameters. As the sound horizon is reduced, $H_0$ increases monotonically, accompanied by a systematic increase in $w_a$, while $w_0$ exhibits only a mild drift. This behavior reflects the geometric nature of BAO constraints, which fix the ratios $D_M(z)/r_d$ and $H(z) r_d$, so that lowering $r_d$ requires a higher late-time expansion rate to preserve consistency with the observed distance measures. As a consequence, the inferred CPL trajectory moves continuously away from the phantom regime, reducing the degree of apparent dark-energy dynamics required by the data. This demonstrates that the shift toward less dynamical, quintessence-like behavior is generic and monotonic, rather than the result of tuning a specific sound-horizon calibration.

We also confirmed that $H_0$ data does not change the results, as shown in the right panel of Fig.~\ref{fig:triangle} (dashed yellow), showcasing the tight relation between $r_d$ and $H_0$.

Finally, in Fig.~\ref{fig:wz_binned}, we compare the CPL results with a binning reconstruction approach (similar to Fig.~12 in~\cite{DESI:2025zgx}, using the formalism described in~\cite{DESI:2025fii}), using BAO+BBN+SN+rs data. We find that the prior on alower $r_d$ leads to a significant shift in the reconstructed $w(z)$, pushing it away from phantom DE ($w<-1$), with the lowest-redshift bin fitting remarkably well to the CPL prediction.
However, while the CPL prediction is consistent with the second bin at the $1\sigma$ level, it departs significantly in the third, suggesting that the $w_0$–$w_a$ parametrization is insufficient to capture the full evolution.
The last bin ($z>2.1$), containing constraints from Ly$\alpha$, is completely unconstrained, except for the top edge, where the posterior steeply goes to zero. This is expected, as values of $w(z)\sim 0$ are non-negligible at early times and are therefore disfavored to maintain consistency with the imposed $r_d$ prior.

Several limitations qualify our null test interpretation. First, our phenomenological approach—imposing a lower sound horizon without altering early-universe physics—neglects model-dependent effects that could influence the late-time parameter inference, as seen in the impact of BBN constraints. Second, the CPL parametrization may not capture the full range of possible DE dynamics, as could be indicated from Fig.~\ref{fig:wz_binned}, particularly if the true equation of state exhibits non-monotonic or rapid variation~\cite{Nesseris:2025lke,Wolf:2023uno,Wolf:2025jlc,Wang:2025znm}. Third, we do not account for potential systematics in the datasets, which could bias our inferences (e.g., \cite{Cortes:2025joz}). Lastly, the exclusion of CMB anisotropy data means we cannot fully assess consistency with the standard cosmological model.

\section{Conclusion}

We presented a model-independent null test to assess how a phenomenologically reduced sound horizon—typically invoked by early-time solutions to the Hubble tension—affects late-time DE inference. Using only low-redshift observables and BBN, we find that a shorter $r_d$ drives the equation of state toward less dynamical, quintessence-like behavior, broadly consistent with $\Lambda$CDM. This highlights the importance of carefully interpreting DE constraints in the context of early-universe assumptions. In particular, some of the apparent late-time evolution in the equation of state may reflect a calibration effect tied to the sound horizon, rather than genuine DE dynamics.

\begin{acknowledgments}
We thank Kris Pardo and Ely Kovetz for insightful comments and valuable discussions. TA is supported in part by the Zuckerman STEM Leadership Program.
\end{acknowledgments}

\bibliography{refs.bib}

\end{document}